\documentclass[twocolumn,twoside,preprintnumbers,supercriptaddress,nofootinbib]{revtex4}

\usepackage{graphicx}% Include figure files
\usepackage{bm}% bold math
\usepackage{fancyhdr}
\usepackage{amsmath,amssymb}
\usepackage{pslatex}

\newcommand{\vk}{\mathbf{k}}
\newcommand{\vq}{\mathbf{q}}
\newcommand{\vx}{\mathbf{x}}
\newcommand{\vy}{\mathbf{y}}
\newcommand{\vz}{\mathbf{z}}
\newcommand{\vu}{\mathbf{u}}
\newcommand{\vv}{\mathbf{v}}
\newcommand{\vvr}{\mathbf{r}}
\newcommand{\vb}{\mathbf{b}}

\newcommand{\Tt}{\tilde{T}}

\newcommand{\abar}{\bar\alpha}

\newcommand{\avg}[1]{\left\langle#1\right\rangle}

\begin{document}

\title{\Large Saturation in High-energy QCD}

\author{Gr\'egory Soyez\footnote{On leave from the fundamental theoretical physics group of the University of Li\`ege.}}
\affiliation{SPhT, CEA/Saclay, Orme des Merisiers, B\^at 774, F-91191 Gif-sur-Yvette cedex, France}
\email{gsoyez@spht.saclay.cea.fr}
%\received{} % It is always \today, today,

\begin{abstract}
In these proceedings, I shall review the basic concepts of perturbative QCD in its high-energy limit, emphasising the approach to the unitarity limit, usually referred to as {\em saturation}. I shall explain the basic framework showing the need for saturation, first, from a simple picture of the high-energy behaviour, then, giving a short derivation of the equation driving this evolution. In the second part, I shall exhibit an analogy with statistical physics and show how this allows to derive {\em geometric scaling} in QCD with saturation. I shall finally consider the effects of gluon-number fluctuations on this picture.
\end{abstract}

\maketitle
\thispagestyle{fancy}

\section{Introduction}

The quest for the high-energy behaviour of perturbative QCD started thirty years ago, soon after QCD was proposed as the fundamental theory of strong interactions. To grasp this problem, it has been realised that saturation effects were to be considered in order to satisfy unitarity constraints. Throughout these proceedings, I shall summarised the present status of our understanding of this limit, which basically means finding an equation giving the evolution of amplitudes towards high energy and finding the properties of its solutions.

\begin{figure}[htbp]
\centerline{\includegraphics[width=8.4cm]{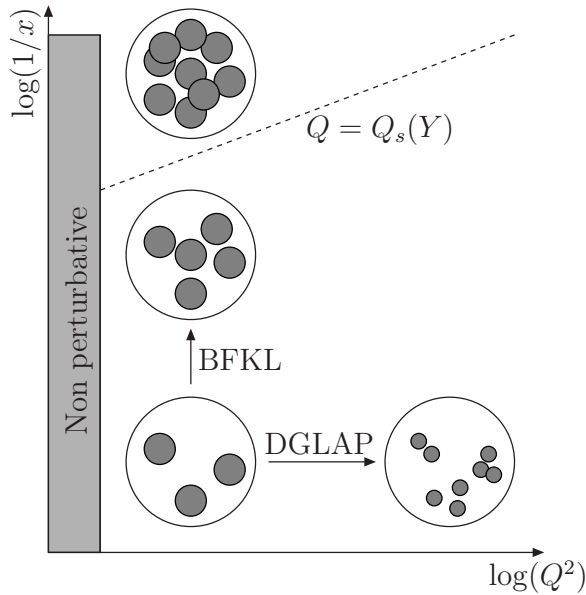}}
\caption{Picture of the proton in DIS}\label{fig:dis}
\end{figure}

In order to motivate the physical picture we want to reproduce, let us consider the problem of Deep Inelastic Scattering (DIS) {\em i.e.} $\gamma^*p\to X$ collisions. This process depends on two kinematical variables: the virtuality $Q^2$ of the photon, measuring the resolution at which the photon scans the proton, and the Bjorken $x$, given by the fraction of the proton momentum carried by the parton struck by the virtual photon in a frame where the proton is moving fast. This last variable is related to the centre-of-mass energy $s=Q^2/x$, meaning that the high-energy limit corresponds to the small-$x$ limit. For practical purposes, we introduce the {\em rapidity} defined through $Y=\log(1/x)$. 

In Figure \ref{fig:dis}, we have represented the typical configuration of the proton in different domains of this phase space. We start with a proton at low $Q^2$ and low energy, represented as three partons (bottom-left part of the picture). If we increase the virtuality of the photon, we are able to resolve its partons into smaller ones. This type of evolution is described in QCD by the Dokshitzer-Gribov-Lipatov-Altarelli-Parisi (DGLAP) equation \cite{dglap}. In that limit, the number of partons rises logarithmically while their typical size decreases like $1/Q^2$ so that the proton becomes more and more dilute. In what follows, we shall be interested in the other type of evolution, namely the evolution towards smaller values of $x$ for constant $Q^2$. If one boosts the proton, what typically happens is that we create new partons of a size comparable to the parents' ones. Obviously, during this evolution, known as the Balitsky-Fadin-Kuraev-Lipatov (BFKL) evolution \cite{bfkl}, the proton tends to the black-disc limit for which partons start to overlap. At that level, they start to interact among themselves and unitarity corrections are to be considered \cite{glr}. This blackening of the proton is called saturation and is what we shall describe from perturbative QCD in these proceedings.

We shall begin our analysis with a description of the BFKL equation. This naturally leads to the introduction of unitarity corrections and to new evolution equations - the Balitsky/JIMWLK hierarchy \cite{balitsky,jimwlk} and the Balitsky-Kovchegov (BK) equation \cite{balitsky,kovchegov}. We shall then show that, within relevant approximations, the BK equation reduces \cite{mp} to the Fisher-Kolmogorov-Petrovsky-Piscounov (F-KPP) equation \cite{fkpp}, well studied in statistical physics. This allows for a description \cite{mp} of the asymptotic behaviour of the scattering amplitudes in terms of travelling waves, translating in QCD into {\em geometric scaling } \cite{geomscaling}. In the last section, we shall discuss the effect of gluon-number fluctuations, which have recently proven \cite{ms,imm,it,msw} to have important consequences on the approach to saturation.

\section{Evolution towards high energy}

Let us start with a fast-moving quark of momentum $p$. This quark emits Bremsstrahlung gluons characterised by their transverse momentum $\vk_\perp^2$ and their longitudinal momentum $k_z=x p$. The probability for emitting such a gluon is
\[
dP \sim \alpha_s \frac{dx}{x}\,\frac{d\vk_\perp^2}{\vk_\perp^2}.
\]
When we consider large transverse momentum (high $Q^2$ in DIS), the collinear divergences $d\vk_\perp^2/\vk_\perp^2$ need to be resummed and this gives rise to a renormalisation group equation known as the DGLAP equation. Throughout this proceedings, we shall instead consider the limit of high energy, in which, we are sensitive to the gluons of small longitudinal momentum $x\ll 1$. This leaves a large phase-space to have successive gluon emissions satisfying $1 \gg x_1 \gg \dots \gg x_n \gg x$ which gives a contribution of order
\[
\alpha_s^n \int_x^1\frac{dx_1}{x_1}\,\dots\,\int_{x_{n-1}}^1\frac{dx_n}{x_n} = \frac{1}{n!}\alpha^n \log(1/x)^n.
\]
Even when the coupling $\alpha_s$ is small enough to ensure applicability of perturbation theory, for small values of $x$ {\em i.e.} high-energy, $\alpha_s\log(1/x)\sim 1$ and all these contributions have to be resummed. 

\begin{figure}[htbp]
\centerline{\includegraphics[width=8.4cm]{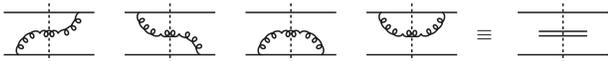}}
\caption{Gluon emission in the dipole picture}\label{fig:dip_split}
\end{figure}

In order to simplify the discussion, let us consider the case of a large number of colours. In that limit, a gluon of transverse coordinate $\vz$ can be considered as a quark-antiquark pair at point $\vz$. In this framework, it is convenient to consider the evolution of a colourless quark-antiquark dipole \cite{mueller}. Indeed, at high-energy, all possible gluon emissions from one dipole are tantamount to this dipole splitting into two child dipoles as depicted in figure \ref{fig:dip_split}. We can thus forget about gluons themselves and talk only in term of dipoles.

Let us denotes by $\avg{T_{\vx\vy}}$ the scattering amplitude\footnote{The notation $\avg{\cdot}$ denotes the average over all possible realisations of the target colour field. It can be understood as the expectation value for the $T$-matrix operator over the target wavefunction.} for a dipole made of a quark at transverse coordinate $\vx$ and an antiquark at $\vy$. Getting the evolution of $\avg{T}$ from the dipole picture is pretty straightforward. Indeed, if one boosts the dipole $(\vx,\vy)$ from a rapidity $Y$ to a rapidity $Y+\delta Y$, this dipole shall split into two dipoles $(\vx,\vz)$ and $(\vz,\vy)$. Each of those dipoles can interact with the target which leads to the following evolution equation\footnote{The last term, coming with a minus sign, corresponds to virtual corrections} for $\avg{T}$
\begin{equation}\label{eq:bfkl}
\partial_Y\avg{T_{\vx\vy}} = \frac{\abar}{2\pi} \int_z {\cal M}_{\vx\vy\vz}
                              \left(\avg{T_{\vx\vz}}+\avg{T_{\vz\vy}}-\avg{T_{\vx\vy}}\right)
\end{equation}
with $\abar=\alpha_s N_c/\pi$ and 
\begin{eqnarray*}
{\cal M}_{\vx\vy\vz} 
& = & \left[\frac{\vx-\vz}{(\vx-\vz)^2}-\frac{\vy-\vz}{(\vy-\vz)^2}\right]^2\\
& = &\frac{(\vx-\vy)^2}{(\vx-\vz)^2(\vz-\vy)^2}
\end{eqnarray*}
is the probability density for dipole splitting\footnote{The first line shows explicitly the two contributions coming from emission by the quark and the antiquark as depicted in figure \ref{fig:dip_split}.}. Equation \eqref{eq:bfkl} is the BFKL equation, derived in the mid-seventies \cite{bfkl} by Balitsky, Fadin, Kuraev and Lipatov. 

This equation being linear, one expects its solution to grow exponentially. To be more explicit, if one neglects impact-parameter dependence, {\em i.e.} assumes that $\avg{T_{\vx\vy}}=\avg{T(r=|\vx-\vy|)}$, the solution of \eqref{eq:bfkl} is
\begin{equation}\label{eq:solbfkl}
\avg{T(r)} = \int \frac{d\gamma}{2i\pi}\,T_0(\gamma)\,e^{\chi(\gamma)Y-\gamma \log(r_0^2/r^2)},
\end{equation}
where $\chi(\gamma)=2 \psi(1)-\psi(\gamma)-\psi(1-\gamma)$ and $T_0(\gamma)$ describes the initial condition. For a fixed dipole size, one finds the high energy behaviour in the saddle point approximation
\[
\avg{T(r)} \sim \exp\left[\omega_P Y-\frac{\log^2(r_0^2/r)}{2\chi''({\scriptstyle{1/2}})\abar Y} \right],
\]
with $\omega_P=4\pi\log(2)\abar$ being the BFKL pomeron intercept. This expression suffers from two major problems: first, even if one start with an initial condition peaked around a small value of $r$, when rapidity increases, the amplitude starts to diffuse to the large dipole sizes {\em i.e.} to the non-perturbative domain. Second, the solution of the BFKL equation grows exponentially with rapidity. It hence violates the unitarity constraint $T\le 1$ obtained from first principles.

\begin{figure}
\includegraphics{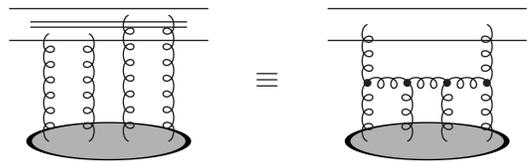}
\caption{Multiple scattering between a target and a projectile made of two dipoles. This can be interpreted as pomeron merging in the target.}\label{fig:multiple}
\end{figure}

This unitarity problem is precisely the point where we meet the requirement to take into account multiple interactions. The most straightforward way to see this is to come back to the splitting of a dipole $(\vx,\vy)$ into $(\vx,\vz)$ and $(\vz,\vy)$. When the target becomes dense enough, both dipoles can scatter on it. We thus have to take into account this contribution, represented in figure \ref{fig:multiple}. This leads to a quadratic suppression term in the evolution equation which becomes
\begin{equation}\label{eq:balitsky}
\partial_Y\avg{T_{\vx\vy}} = \frac{\abar}{2\pi} \int_z {\cal M}_{\vx\vy\vz}
                              \left(\avg{T_{\vx\vz}}+\avg{T_{\vz\vy}}-\avg{T_{\vx\vy}}-\avg{T^2_{\vx\vz;\vz\vy}}\right).
\end{equation}
The first thing to remark is that this new term is of the same order as the previous ones when $T^2\sim T$ or, equivalently, when $T\sim 1$. It is thus, as expected, a mandatory contribution near the unitarity limit. However, equation \eqref{eq:balitsky} involves a new object, namely $\avg{T^2_{\vx\vz;\vz\vy}}$ which probes correlations inside of the target. In a general framework, one should then write down an equation for $\avg{T^2}$, which will involve $\avg{T^2}$ through BFKL-like contributions and $\avg{T^3}$ from unitarity requirements. This ends up with a complete hierarchy, giving the evolution for each $\avg{T^k}$, known as the (large-$N_c$) Balitsky hierarchy \cite{balitsky}.

If the target is sufficiently large and homogeneous, one can simply assume $\avg{T^2_{\vx\vz;\vz\vy}}= \avg{T_{\vx\vz}}\avg{T_{\vz\vy}}$, ending up with a closed equation for $\avg{T}$
\begin{equation}\label{eq:bk}
\partial_Y\avg{T_{\vx\vy}} = \frac{\abar}{2\pi} \int_z {\cal M}_{\vx\vy\vz}
                             \left(\avg{T_{\vx\vz}}+\avg{T_{\vz\vy}}-\avg{T_{\vx\vy}}-\avg{T_{\vx\vz}}\avg{T_{\vz\vy}}
                             \right).
\end{equation}

This last expression is the Balitsky-Kovchegov (BK) equation \cite{balitsky, kovchegov}, which is the most simple equation one can obtain from perturbative QCD by including both the BFKL contributions at high-energy and the corrections from unitarity. It is easy to check that $\avg{T}=0$ is an unstable fix point of the BK equation, while $\avg{T}=1$ is a stable fix point, ensuring unitarity is satisfied. In general, unitarity effects cut the increase of the amplitude for scales smaller than a typical scale, called the {\em saturation momentum}, increasing with rapidity. In particular, it solves the problem of diffusion to the infrared obtained with linear BFKL equation. We shall come back with more details to the study of the solution of the BK equation for the next section.

If one wants to study those effects beyond the large-$N_c$ approximation, one better used the colour-glass-condensate (CGC) formalism. This is an effective theory in which one consider the radiation of soft gluons in a strong background field which introduces the nonlinear effects requested at saturation. At a given rapidity, the target is considered as a random colour source $\rho^a(\vx)$ and radiates a field according to the QCD Yang-Mills equations. We then study the rapidity evolution of the functional $W_Y[\rho]$ giving the probability to find a source $\rho$ at rapidity $Y$. This functional allow to compute the average value of an operator using
\begin{equation}\label{eq:cgcavg}
\avg{{\cal O}}_Y = \int {\cal D}[\rho]\:W_Y[\rho]\,O[\rho].
\end{equation}

\begin{figure}
\includegraphics[scale=0.9]{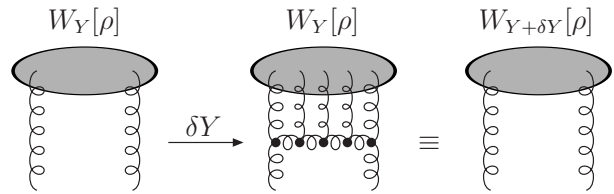}
\caption{One step of JIMWLK evolution: the emitted soft gluon propagating in the string target field is used to redefine the probability $W[ \rho]$ at rapidity $Y+\delta Y$. The case of two gluons only corresponds to BFKL.}\label{fig:cgc}
\end{figure}

We thus need to find the evolution of $W[\rho]$. One step of evolution is represented diagrammatically in figure \ref{fig:cgc}. As for the BFKL equation, one soft gluon is radiated (yielding a $\alpha_s \log(1/x)$ factor). The difference comes from the fact that when the target is sufficiently dense, one has to dress the propagator of this soft gluon with the field created by the target. These emissions are then reabsorbed inside the probability $W[\rho]$ in order to define it at rapidity $Y+\delta Y$. Each background gluon insertion gives a factor $g\alpha$ with $g$ the QCD coupling and $\alpha$ the $A^+$ colour field of the target. Including these effects becomes mandatory when the field is of order $1/g$. The amplitude is defined through 
\begin{equation}\label{eq:cgct}
\avg{T_{\vx\vy}}=1-\frac{1}{N_c}\avg{{\rm tr}\left(W_\vx^\dagger W_\vy\right)},
\end{equation}
where $W_\vx$ is the Wilson line operator
\[
W_\vx = P\:\exp\left[ig \int_{-\infty}^\infty dx^- \alpha(x^-,\vx) \right],
\]
which means that in the leading order, $\avg{T_{\vx\vy}}\propto g^2(\alpha_\vx-\alpha_\vy)^2$. This shows that the strong field $\alpha\sim 1/g$ corresponds to $T\sim 1$ and thus to saturation corrections. A careful treatment of the diagrams in figure \ref{fig:cgc} leads to the JIMWLK equation \cite{jimwlk}
\begin{equation}\label{eq:jimwlk}
\partial_Y W_Y[\rho]
 = \int {\cal D}[\rho]\,\frac{\delta}{\delta \rho^a_\vx}\chi^{ab}_{\vx\vy}\frac{\delta}{\delta \rho^b_\vy} W_Y[\rho],
\end{equation}
with the kernel $\chi$ computable in perturbative QCD.

\begin{figure*}
\includegraphics{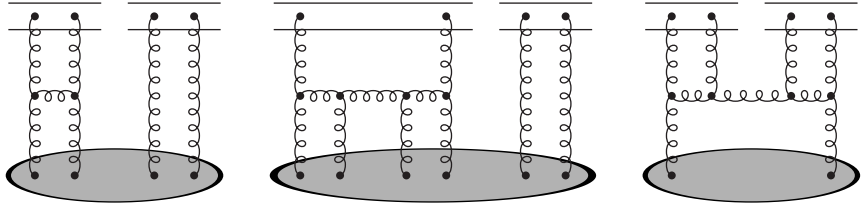}
\caption{Diagrams contributing to the evolution of $\avg{T^2}$. They correspond, from left to right, to BFKL ladders, saturation corrections and fluctuations effects.}\label{fig:t2}
\end{figure*}

This functional integro-differential equation gives the rapidity evolution of $W[\rho]$. Using relations like \eqref{eq:cgcavg} or \eqref{eq:cgct}, one can obtain the evolution of the scattering amplitudes. The resulting equations are equivalent to the Balitsky hierarchy \cite{balitsky}. These equations going beyond large $N_c$, they involves non-dipolar operators. For example, the equation for two dipoles depends on a sextupole operator (build out of six Wilson lines). The formalism beyond large $N_c$ being rather tricky, in what follows we shall stick to the BK equation, which is sufficient to grasp the basic concepts of saturation.

Very recently, it has been realised that the question of the high-energy limit of QCD, even at the leading logarithmic accuracy, is not fully described by the Balitsky hierarchy. To see this more precisely, let us consider in more details the evolution equation for $\avg{T^2}$. It corresponds to the scattering of two dipoles off a target and the diagrams contributing to one step of rapidity evolution are shown in figure \ref{fig:t2}. The first graph on the left of this figure is simply the BFKL-ladder contribution for which $\partial_Y\avg{T^2}\propto \avg{T^2}$. The diagram in the middle corresponds to multiple scattering (see also figure \ref{fig:multiple}). It accounts for unitarity corrections and gives a contribution $\partial_Y\avg{T^2}\propto \avg{T^3}$. If those two graphs were the only relevant ones in the evolution, one would obtain the (large-$N_c$) Balitsky equation. Nonetheless, it has been argued that a third contribution, represented by the rightmost diagram of figure \ref{fig:t2}, has to be included. This graph describes multiple scattering off the projectile or, equivalently, gluon-number fluctuations inside the target. After the pomeron-merging term, contributing to saturation, one thus now also includes the pomeron-splitting contribution, which gives rise to a new term in the evolution equation $\partial_Y\avg{T^2}\propto \avg{T}$. For large-$N_c$ and at the two-gluon-exchange level, this new term has been computed \cite{it} and leads to
\begin{eqnarray}\label{eq:fluct}
\lefteqn{\left.\partial_Y\avg{T^2_{\vx_1\vy_1;\vx_2\vy_2}}\right|_{\text{fluct}}
 = \frac{1}{2}\frac{\abar}{2\pi}\left(\frac{\alpha_s}{2\pi}\right)^2}\\
&& \int_{\vu\vv\vz} {\cal M}_{\vu\vv\vz} {\cal A}_0(\vx_1\vy_1|\vu\vz){\cal A}_0(\vx_2\vy_2|\vz\vv) \nonumber
   \nabla_\vu^2 \nabla_\vv^2 \avg{T_{\vu\vv}} + (1 \leftrightarrow 2),
\end{eqnarray}
with ${\cal A}_0$ being the dipole-dipole amplitude at the two-gluon-exchange level
\[
{\cal A}_0(\vx\vy|\vu\vv) = \frac{1}{8}\log^2\left[\frac{(\vx-\vu)^2(\vy-\vv)^2}{(\vx-\vv)^2(\vy-\vu)^2} \right].
\]
This object naturally comes out when we relate the scattering amplitude with the dipole density $n$:
\begin{equation}\label{eq:ntoT}
\avg{T_{\vx\vy}} = \alpha_s^2\int_{\vu\vv} {\cal A}_0(\vx\vy|\vu\vv)\,\avg{n_{\vu\vv}}.
\end{equation}
This relation can be inverted to obtain $\avg{n_{\vx\vy}}=\alpha_s^{-2}\nabla_\vx^2 \nabla_\vy^2\avg{T_{\vx\vy}}$.

The correction \eqref{eq:fluct} becomes of the same order as the BFKL contribution when $T^2\sim \alpha_s^2 T$, or $T\sim \alpha_s^2$, {\em i.e.} in the dilute regime where, as expected, fluctuations should lead to important effects. 
Although this may at first sight seems irrelevant for the physics of saturation, one has to realise that, due to colour transparency, the amplitude becomes arbitrarily small for small dipole size, hence, even for dense objects, there is always some region of the phase-space where the system is dilute and its evolution governed by fluctuation effects. Once the system has evolved through a fluctuation, it grows, from BFKL evolution, like two pomerons, {\em i.e.} $T\sim\alpha_s^2\exp(2\omega_P Y)$. This has to be compared with the one-pomeron exchange $T\sim \exp(\omega_P Y)$. We see that at very large energies $Y\ge \omega_P^{-1}\log(1/\alpha_s^2)$, BFKL evolution compensates the extra factor of $\alpha_s^2$ coming from the initial fluctuation. We shall see more precisely in the last section that those fluctuation effects in the dilute tail have important consequence on the saturation physics.

Finally, let us conclude this section by quoting that the equation \eqref{eq:fluct}, including linear, saturation and fluctuation effects is the most complete equation known in perturbative high-energy QCD so far. Its extension beyond the large-$N_c$ approximation is still a challenging problem.

\section{Properties of the scattering amplitudes}

In this section, we shall sketch out the main properties of the solutions of the evolution equations towards high energy. As we shall explain in detail in the next lines the major source of information comes from the analogy between the BK equation and the Fisher-Kolmogorov-Petrovsky-Piscounov (F-KPP) equation which is very well studied in statistical physics. This section shall be devoted to derive this analogy and its consequences in the case of the BK equation. We leave the discussion concerning the effect of fluctuations for the next section.

So, let us start with the BK equation \eqref{eq:bk}. We shall first restrict ourselves to the impact-parameter-independent version of the equation, for which one can easily go to momentum space by using\footnote{In this section, we omit the $\avg{\cdot}$ quotations as they do not play any role in the mean-field approximation.}
\[
\Tt(k) = \frac{1}{2\pi}\int \frac{d^2r}{r^2}\,e^{i\vk.\vvr}\,T(r) 
       = \int_0^\infty \frac{dr}{r}\, J_0(kr)\, T(r),
\]
and where the equation becomes
\begin{equation}\label{eq:bkk}
\partial_Y \Tt(k) = \abar \chi(-\partial_L)\Tt(k) - \abar \Tt^2(k),
\end{equation}
with $L = \log(k^2/k_0^2)$, $k_0$ being a soft reference scale. 

In order to simplify the problem, we shall work in the saddle point approximation (often referred to as the {\em diffusive approximation}) and expand the BFKL kernel to second order around $\gamma=1/2$, transforming the complicated differential operator in equation \eqref{eq:bkk} into a second-order operator
\[
\chi(-\partial_L)\Tt(k) \approx \chi({\scriptstyle \frac{1}{2}}) 
 + {\scriptstyle \frac{1}{2}}\chi''({\scriptstyle \frac{1}{2}})\left(\partial_L+{\scriptstyle \frac{1}{2}}\right)^2.
\]
Up to a linear change of variable switching from $Y$ and $L$ to time $t=\abar Y$ and space $x = L + \text{cst}.Y$ and a renormalisation of the amplitude $T\to u=\text{cst}.T$, the BK equation in the saddle point approximation becomes
\[
\partial_t u(x,t) = \partial_x^2u(x,t) + u(x,t) - u^2(x,t).
\]
This equation is nothing but the F-KPP equation studied in statistical physics since forty years and applying to many problem. It describes reaction diffusion processes in the mean-field approximation where one can have creation and annihilation of particles locally and diffusion to neighbouring site {\em e.g.} the density of population in the centre of Rio.

\begin{figure}
\includegraphics[scale=0.8]{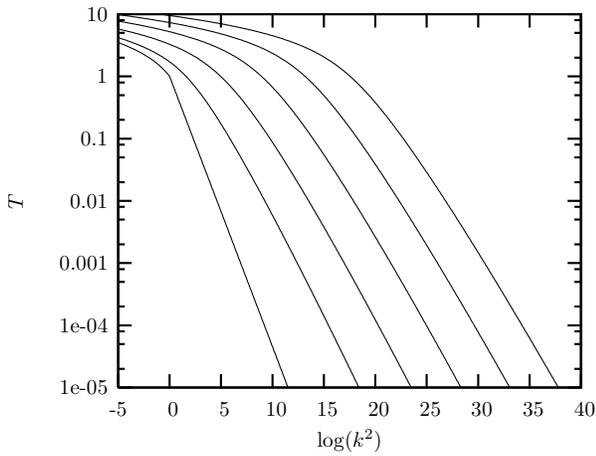}
\caption{Numerical simulation of the rapidity-evolution of the BK equation. We start at $Y=0$ from the leftmost amplitude and evolve to higher $Y$ using \eqref{eq:bkk}. Amplitude is shown for $Y=5,10,15,20,25$ and clearly exhibits a travelling-wave pattern.}\label{fig:bkfront}
\end{figure}

\begin{figure}
\includegraphics[scale=0.8]{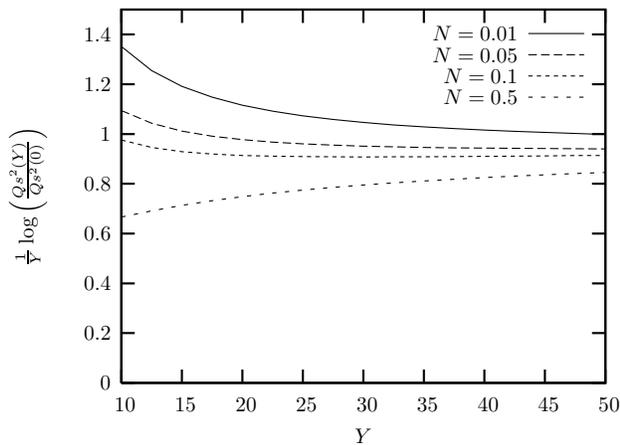}
\caption{Rapidity evolution of the saturation scale extracted from figure \ref{fig:bkfront}. More precisely, the speed of the wave, {\em i.e.} the exponent of the saturation scale, is plotted. It goes to a constant as it should.}\label{fig:bkqs}
\end{figure}

One knows that the F-KPP equation admits travelling waves as asymptotic solutions {\em i.e.} there exists a critical velocity $v_c$ and a critical slope $\gamma_c$, determined only from the knowledge of the linear kernel of the F-KPP equation $\partial_x^2+\mathbf{1}$, such that (see figure \ref{fig:bkfront} for a pictorial representation coming from the BK equation which we shall explain a bit later)
\[
u(x,t) \stackrel{t\to \infty}{\longrightarrow} u(x-v_ct) \stackrel{x\gg v_ct}{\approx} e^{-\gamma_c(x-v_ct)} \log(x-v_ct).
\]
Although working in the diffusive approximation makes the discussion easier due to the direct analogy with the F-KPP equation, this assumption is not required. Indeed, for an equation with a more complicated kernel (than $\partial_x^2+\mathbf{1}$), one can prove the existence of travelling waves provided the three following conditions are satisfied
\begin{enumerate}
\item the amplitude has $0$ as unstable fix point and $1$ as stable fix point,
\item the initial condition decreases faster than $\exp(-\gamma_c L)$ at large $L$ (see below for the general definition of $\gamma_c$),
\item the equation obtained by neglecting the nonlinear terms admits superposition of waves as solution.
\end{enumerate}
Then, travelling waves are formed during evolution with the critical parameters obtained, from the linear kernel only, through the relation
\[
v_c=\chi'(\gamma_c)=\frac{\chi(\gamma_c)}{\gamma_c}.
\]
One can show that this corresponds to the selection of the wave with the minimal speed $\chi(\gamma)/\gamma$ and the point where group and phase velocities are equal.

In the case of the BK equation, the first condition is satisfied due to BFKL growth and saturation, the second comes from colour transparency ($T\propto 1/k^2=\exp(-L)$) and the last one is equivalent to the solution \eqref{eq:solbfkl} (with $r/r_0$ replaced by $k_0/k$) of the BFKL equation obtained by neglecting nonlinear terms in \eqref{eq:bkk}.

The critical parameters (for the complete BFKL kernel {\em i.e.} for \eqref{eq:bkk}), are found to be $\gamma_c\approx 0.6275$ and $v_c\approx 4.8836\abar$. Written in terms of the QCD variables $Y$ and $k^2$, one can then write the asymptotic solution for the impact-parameter-independent BK equation under the form
\begin{equation}\label{eq:bkfront}
T(k;Y)
 \stackrel{Y\to\infty}{=} T\left(\frac{k^2}{Q_s^2(Y)}\right)
 \stackrel{k\gg Q_s}{=}\left[\frac{k^2}{Q_s^2(Y)}\right]^{-\gamma_c}\log\left[\frac{k^2}{Q_s^2(Y)}\right],
\end{equation}
with the saturation scale given by
\begin{equation}\label{eq:bkqs}
Q_s^2(Y) \stackrel{Y\to\infty}{=} k_0^2\,\exp\left[ v_c Y - \frac{3}{2\gamma_c}\log(Y) \right].
\end{equation}

The formation of a travelling-wave pattern when energy increases can easily be seen on numerical simulations of the BK equation. As displayed in figure \ref{fig:bkfront}, if one start with a steep enough initial condition (leftmost curve), the amplitude increases with energy and a wave moving into the dilute domain gets formed. From that simulation, one can extract the saturation scale by solving $T(Q_s^2,Y)=N$ at each values of $Y$ and for a fixed threshold $N$. In figure \ref{fig:bkqs}, we have plotted $\log(Q_s^2)/Y$ which goes to a constant value as it is expected from \eqref{eq:bkqs}.

Equation \eqref{eq:bkfront} has a remarkable property: it proves that at high energy the amplitude, {\em a priori} a function of both $Y$ and $k$, depends only on the ratio between $k$ and the saturation momentum. This property, known as {\em geometric scaling}, has been observed \cite{geomscaling} in the HERA measurements of the proton structure function. The fact that geometric scaling can be derived from the BK equation is one of the most important indication for the experimental observation of saturation. The saturation scale obtained from the structure function data is of the order of 1 GeV for $x\sim 10^{-5}$. This means that high-energy QCD can indeed be studied from perturbation theory.

Up to now, we have only discussed the impact-parameter independent BK equation. We might therefore ask whether or not these arguments extend to the full equation including all phase-space degrees of freedom. A dipole of transverse coordinates $(\vx,\vy)$ is then represented through its size $\vvr = \vx-\vy$ and impact parameter $\vb=(\vx+\vy)/2$. The problem is then that dipole splitting is non-local in impact parameter. This means that the BK equation couples different values of $\vb$ and we can not apply directly the previous arguments for each value of $\vb$. Again, the solution consists in moving to momentum space and replace the impact parameter by momentum transfer $\vq$
\[
\tilde T(\vk,\vq) = \int d^2x\,d^2y\,e^{i\vk.\vx}e^{i(\vq-\vk).\vy}\frac{T(\vx,\vy)}{(\vx-\vy)^2}.
\]
The BK equation then takes a form \cite{bdep} for which the BFKL kernel is local in $\vq$
\begin{eqnarray}\label{eq:bkfull}
\lefteqn{\partial_Y \tilde T(\vk,\vq)}\\
& = & \frac{\bar\alpha}{\pi}\int \frac{d^2k'}{(k-k')^2}\left\{
      \tilde T(\vk',\vq) - \frac{1}{4}\left\lbrack
      \frac{k^2}{k'^2}+\frac{(q-k)^2}{(q-k')^2}
      \right\rbrack\tilde T(\vk,\vq)\right\}\nonumber\\
& - & \frac{\bar\alpha}{2\pi}\int d^2k'\,\tilde T(\vk,\vk')\tilde T(\vk-\vk',\vq-\vk').\nonumber
\end{eqnarray}

We can now proceed with this equation in a similar way as for the the impact-parameter-independent case. The existence of travelling waves required the three conditions stated previously to be satisfied. The two first ones are trivially satisfied \footnote{Indeed, we have saturation and BFKL growth ensuring the first condition and colour transparency still apply for the second one.}. For the third condition to be valid, we need to find solutions of the linear part of \eqref{eq:bkfull}, {\em i.e.} the complete BFKL equation, which can be expressed as a superposition of waves. This is merely technical as it consists in a careful treatment of the solutions of the BFKL equation \cite{bfklsol} in its full glory, so I shall directly jump to the results. We can show \cite{bdep2} that when the dipole momentum $k$ is much larger than the momentum transfer $q$ and the typical scale of the target $k_0$, the solutions of the BFKL equation are a superposition of waves and therefore, we obtain travelling waves for the full BK equation. More precisely, the high-energy behaviour of the amplitude takes the form
\begin{eqnarray*}
\tilde T(k,q;Y)
& \stackrel{Y\to\infty}{=} & T\left(\frac{k^2}{Q_s^2(q;Y)}\right)\\
& \stackrel{k\gg Q_s}{=}   & \left[\frac{k^2}{Q_s^2(q;Y)}\right]^{-\gamma_c}\log\left[\frac{k^2}{Q_s^2(q;Y)}\right].
\end{eqnarray*}
This expression is exactly the same as for the previous case except that the saturation scale now depends on momentum transfer
\[
Q_s^2(q;Y) \stackrel{Y\to\infty}{=} \Lambda^2\;\exp\left[v_c Y - \frac{3}{2\gamma_c}\log(Y)\right]
\]
with
\[
\Lambda^2 = \begin{cases}
 k_0^2 & \text{if } k_0 \gg q\\
 q^2   & \text{if } q \gg k_0
\end{cases}.
\]
It is interesting to notice that the critical slope $\gamma_c$ ad speed $v_c$, obtained from the BFKL kernel only, are the same as for the impact-parameter-independent case. 

The important point at this stage is that this study {\em predicts} geometric scaling at non-zero momentum transfer. Experimental measurements of the DVCS cross-sections or diffractive $\rho$-meson production are good candidates and can provide more evidence for saturation.

\section{High-energy QCD with fluctuations}

Let us now consider the effect of adding the fluctuation contribution \eqref{eq:fluct}. This new term, added to the full Balitsky hierarchy, turns a simple non-linear equation into a infinite hierarchy with a complicated transverse-plane dependence\footnote{The fluctuation term involves an awkward integration. By integration by part one can express it as a vertex applied to $T$ but the integration over the internal variable $\vz$ in \eqref{eq:fluct} is unknown.}. We can however simplify the equation to a tractable problem by performing a local-fluctuations approximation. This amounts to simplify the dipole-dipole scattering amplitude ${\cal A}_0$ used to relate the dipole density $n$ with the scattering amplitude $T$. We assume that two dipoles interact if they are of the same size and if their centre-of-mass are sufficiently close to allow for an overlap of the two dipoles. Within this approximation equation \eqref{eq:ntoT} is replaced
\[
\avg{T_{\vx\vy}} = \kappa\alpha_s^2 \,\left|\vx-\vy\right|^4\,\avg{n_{\vx\vy}},
\]
where $\kappa$ is an unknown fudge factor of order 1.

Once this is done, the remaining steps are no particularly illuminating: one Fourier-transform the dipole size $r=|\vx-\vy|$ into momentum $k$ and perform a coarse-graining approximation to get rid of the impact-parameter dependence of the fluctuation term. This computation results into the following simplified form of the hierarchy (we give only the expression for $\avg{T}$ and $\avg{T^2}$ for simplicity)
\begin{eqnarray*}
\partial_Y \avg{T_k} & = & \bar\alpha \chi(-\partial_L)\avg{T_k}
                         - \bar\alpha \avg{T^2_{k,k}}, \\
\partial_Y \avg{T^2_{k_1,k_2}} & = & \bar\alpha \chi(-\partial_{L_1})\avg{T^2_{k_1,k_2}}
                                   - \bar\alpha \avg{T^3_{k_1,k_1,k_2}} + (1 \leftrightarrow 2)\\
            & + & \bar\alpha \, \kappa\alpha_s^2\, k_1^2 \delta(k_1^2-k_2^2)\avg{T_{k_1}},
\end{eqnarray*}
where, as previously, $L_i = \log(k_i^2/k_0^2)$.

Although this form of the equation does not look that much simpler at first sight than the original one, it has the advantage that it can be rewritten under the form of a Langevin equation\footnote{The complete hierarchy using \eqref{eq:fluct} can also be rewritten as a Langevin equation \cite{it}. Unfortunately, its expression is highly nontrivial.}
\begin{equation}\label{eq:langevin}
\partial_Y T(L)
  = \bar\alpha \left[\chi(-\partial_L)T(L)-T^2(L)+\sqrt{\kappa\alpha_s^2T(L)}\eta(L,Y)\right],
\end{equation}
where $\eta$ is a Gaussian white noise satisfying the following commutation relations:
\begin{eqnarray}\label{eq:noise}
\avg{\eta(L,Y)} & = & 0,\nonumber\\[-3mm]
& & \\[-3mm]
\avg{\eta(L_1,Y_1)\eta(L_2,Y_2)} & = & \frac{4}{\bar\alpha}\delta(L_1-L_2)\delta(Y_1-Y_2). \nonumber
\end{eqnarray}

The Langevin equation is a stochastic equation describing an event-by-event picture. Each realisation of the noise term corresponds to a particular evolution of the target and one can show that, once we average over those realisations using the correlations \eqref{eq:noise}, the complete hierarchy for the evolution of $\avg{T^k}$ is recovered. It is interesting to notice that equation \eqref{eq:langevin} is formally equivalent to the BK equation with an additional noise term.

As for the case of the BK equation, one can restrict ourselves to the diffusive approximation. This leads to the stochastic FKPP (sFKPP) equation which amounts to add a noise term to the FKPP equation. It has found many applications in various fields, {\em e.g.} in the description of reaction-diffusion systems. The noise term appears once we have to consider discreteness effects ({\em e.g.} for a reaction-diffusion system on a lattice, only an integer number of particles are admitted per site). When the number of particles involved goes to infinity, the mean-field approximation is justified and the (non-stochastic) FKPP equation is valid. 

\begin{figure}
\includegraphics[scale=0.66]{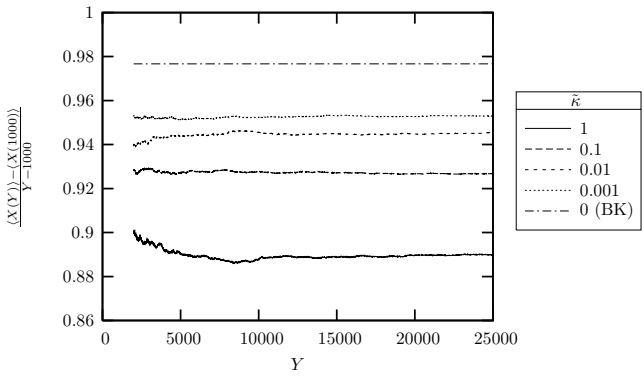}
\caption{Speed of the wave for different values of the noise strength $\kappa\alpha_s^2$. The BK speed has been added for comparison. One sees that the speed decreases when $\kappa\alpha_s^2$ increases.}\label{fig:speed}
\end{figure}

\begin{figure}
\includegraphics[scale=0.71]{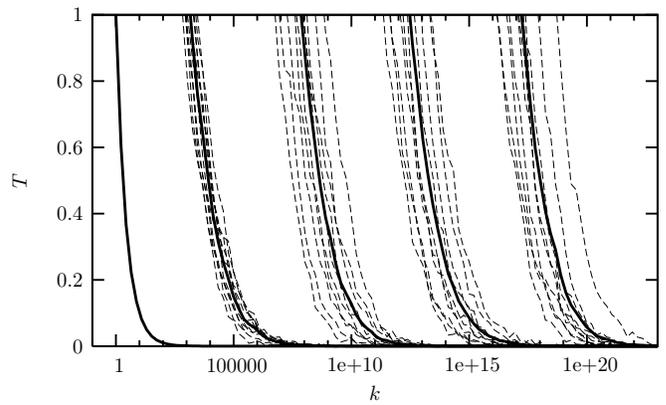}
\caption{Numerical simulations of equation \eqref{eq:langevin}. The dashed curves are different events corresponding to the same initial condition with different realisations of the noise term. Travelling waves are observed for each curve, together with dispersion. The black curve is the average amplitude.}\label{fig:events}
\end{figure}

The basic effects of the additional noise term in the sFKPP equation are known, at least qualitatively. despite the fact that the fluctuations are only expected to have large consequences on the dilute tail of the wavefront, these modifications changes quite a lot the picture at saturation. In order to test the validity of those results in the case beyond the diffusive approximation, {\em i.e.} with the full BFKL kernel, we have performed \cite{gs} numerical studies (see also \cite{egm}) of the QCD equation \eqref{eq:langevin}. Moreover, many analytical results are only known in the limit where the noise strength $\kappa\alpha_s^2$ is (irrealistically) small while we have concentrated our numerical work of physically acceptable values. In the following paragraphs, we review the main effects of the fluctuation contribution and show that they are observed in the numerical analysis.

First of all, for a single event, the evolved amplitude shows a travelling-wave pattern (up to small fluctuations in the far tail which, at least a this level, are irrelevant for the wave pattern at saturation). This means that each single realisation of the noise leads to geometric scaling, as it is the case for the BK equation. At this stage, the major difference comes from a decrease of the speed of the wave. Analytically, this can only be computed in the limit $\kappa\alpha_s^2\to 0$ {\em i.e.} when the fluctuations are a small perturbation around the mean-field behaviour\footnote{A recent analysis has also computed its strong-noise behaviour \cite{strong}.}. It has been found \cite{bd} that 
\[
v^* \underset{\alpha_s^2\kappa\to 0}{\to} v_c - \frac{\abar\pi^2\gamma_c\chi''(\gamma_c)}{2\log^2(\alpha_s^2\kappa)}.
\]
Unfortunately, this expression gives only reliable results for extremely small values of $\kappa\alpha_s^2\to 0$ such as $10^{-20}$. This is clearly not sufficient for realistic situations. The numerical simulations we have performed clearly exhibits this decrease of the speed when the noise strength $\kappa\alpha_s^2$ increases (see figure \ref{fig:speed}).

The second noticeable effect of the noise term is to introduce dispersion between different events. Different events have the same shape but their position $X(Y)$ ($\log(Q_s^2(Y))$ in physical units) fluctuates. At a give rapidity, one can compute the dispersion of those events. This diffusion process being very similar to a random walk, one expects that the dispersion of the events behaves like $\sqrt{Y}$:
\[
\avg{X^2}_Y-\avg{X}_Y^2 \stackrel{Y\to\infty}{\approx} D_{\text{diff}}Y.
\]
The parametric dependence of the diffusion coefficient $D_{\text{diff}}$ has been computed numerically \cite{bd} and behaves\footnote{Recently, this coefficient has been computed analytically \cite{bdmm}.} like $|\log^{-3}(\kappa\alpha_s^2)|$.

\begin{figure}
\includegraphics[scale=0.67]{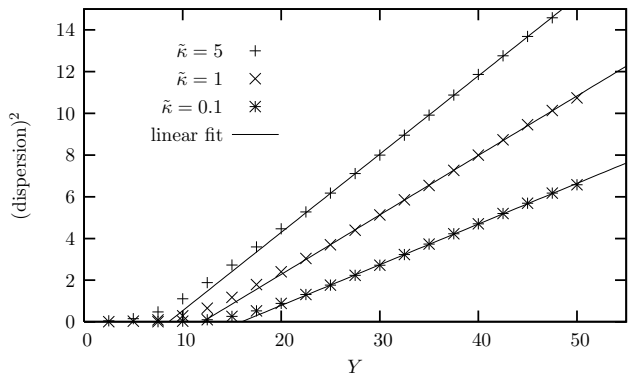}
\caption{Dispersion (squared) of the position of the events as a function of rapidity. As expected, it dispersion increases like $\sqrt{Y}$ but few dispersion is obtained in early stages of the evolution.}\label{fig:disp}
\end{figure}

\begin{figure}
\includegraphics[scale=0.69]{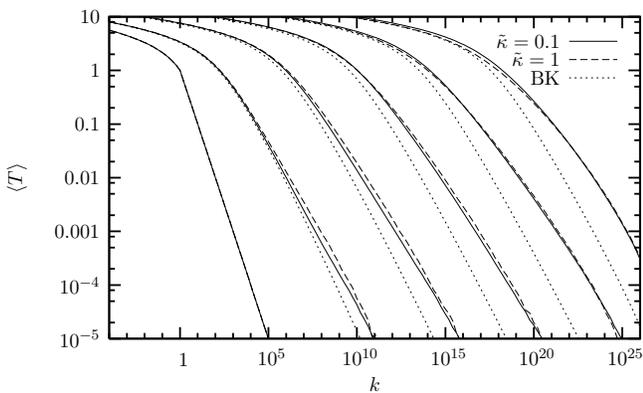}
\caption{Evolution of the averaged amplitude for different values of the noise strength, compared with the mean field result (dotted curve). The curves correspond, from left to right, to $Y=0,10,20,30,40$ and $50$.}\label{fig:front-avg}
\end{figure}

Concerning the numerical simulations, the dispersion of the wavefront between different events and its increase with rapidity are manifest on figure \ref{fig:events}. One can then extract the dispersion as a function of rapidity. The resulting curve, shown on figure \ref{fig:disp} for different values of the noise strength, calls for two remarks. First, the expected asymptotic behaviour (dispersion $\propto \sqrt{Y}$) is observed and the diffusion coefficient increases with rapidity. However, for small values of the rapidity, we do not observe a significant dispersion.

This dispersion of the events has an important physical consequences: although each single event displays geometric scaling, once we compute the average amplitude, dispersion induces geometric scaling violations\footnote{The asymptotic behaviour exhibits a new form of scaling, called {\em diffusive scaling} and its implications on DIS has been recently studied \cite{himst}.}. This effect is best seen on figure \ref{fig:front-avg} where we have compared the evolution with fluctuations to the BK results. While in the BK equation, a fixed travelling-wave pattern is formed, once fluctuations are take into account, we see a broadening of the average amplitude as rapidity increases.

\section{Discussion and perspectives}

Let us now summarise the main points raised in these proceedings. As this document is in itself some kind of a summary, I shall only pick up the main points and refer to various references for detailed approaches.

First, we have shown that it is possible to describe the evolution to high energy in pQCD by a hierarchy of equations (see \eqref{eq:fluct}). This gives the evolution of the $\avg{T}$ matrix (averaged over the target wave-function) and its higher-order correlations $\avg{T^k}$. In this hierarchy, the evolution of $\avg{T^k}$ contains three types of contribution:
\begin{enumerate}
\item the linear BFKL growth, proportional to $\abar\avg{T^k}$. This is the usual high-energy contribution computed thirty years ago and corresponding to the exchange of $k$ pQCD pomerons. It leads to a fast (exponential) increase of the scattering amplitude.
\item the saturation corrections, proportional to $\abar\avg{T^{k+1}}$. This negative term becomes important when the amplitude reaches unitarity and it allows for the constraint $T(\vvr,\vb)\le 1$ to be satisfied.
\item the fluctuations term, proportional to $\abar\alpha_s^2\avg{T^{k-1}}$. These fluctuations correspond to gluon-number fluctuations in the target. They play an important role when the amplitude is of order $\alpha_s^2$ or, equivalently, when the dipole density is of order one, {\em i.e.} in the dilute regime where one indeed expects fluctuation effects to appear.
\end{enumerate}

If one neglects the fluctuation contributions, we recover the Balitsky hierarchy (in its large-$N_c$ formulation) and, in the mean field approximation, the BK equation is obtained. This last equation, much simpler than the infinite hierarchy captures many important points concerning the physics of saturation. We also have to point out that, as long as we do not take into account fluctuations, the evolution at all orders in $1/N_c$ exists under the form of the Balitsky/JIMWLK equation. We have only given a short introduction to the Colour Glass Condensate (in which the JIMWLK equation is derived). Further information concerning the Balitsky (resp. CGC) approach can be found if reference \cite{wilson} (resp. \cite{cgc}).

Many additional things can be said concerning those different contributions, {\em e.g.} concerning the equivalence between descriptions from the target and projectile point of view (see {\em e.g.} \cite{ddd}), or the attempts to describe the evolution beyond its large$-N_c$ limit \cite{beyond}. Again, the interested reader is forwarded to the corresponding references (and the references therein) for further information.

In the second part of this overview, we have concentrated ourselves on the physical consequences arising from saturation and fluctuation effects. We have searched how those contributions manifest themselves on the scattering amplitude and change the exponential behaviour obtained from the BFKL evolution. Again, the most important points can be summarised in two major steps with their respective physical consequences:
\begin{enumerate}
\item Considering BFKL with saturation effects (mean-field picture), one can show that the BK equation lies in the same universality class as the FKPP equation. This implies that travelling waves are formed during the evolution towards high energy. From the BFKL kernel one obtains the critical parameters which are the anomalous dimension observed in the large-$Q^2$ tail and the speed of the wave, equivalent to the exponent of the saturation scale. In physical terms, these travelling waves corresponds to geometric scaling. The experimental observation at HERA can thus be seen as a consequence of saturation. It is extremely important to realise that geometric scaling is a prediction from saturation physics which extends to large values of $Q^2$, far beyond the saturation scale itself\footnote{The geometric scaling window grows, in logarithmic units, like $\sqrt{Y}$ beyond the saturation scale.}. This geometric scaling has also been predicted at nonzero momentum transfer, a result which may be applied for example to $\rho$-meson production or DVCS.
\item If one includes the effects of fluctuations, the evolution becomes a Langevin equation\footnote{If one consider the complete evolution equation with full phase-space dependence, one can also rewrite the hierarchy under the form of a Langevin equation. It is also possible \cite{ist} to view the evolution s a reaction-diffusion process.} equivalent to the stochastic FKPP equation, including a noise term. For each realisation of this noise, we observe geometric scaling with a decrease of the speed w.r.t. the mean-field case. If we consider a bunch of events, we observe that, although they all have the same shape, their position is diffused with diffusion going like $\sqrt{Y}$. This dispersion of the saturation scale for a set of events implies that geometric scaling is violated for the averaged amplitude. Numerical studies indicate that geometric scaling is still valid at small rapidities. At higher energies, a new (diffusive) scaling is predicted \cite{himst} in which the amplitude scales like $\log(k^2/Q_s^2(Y))/\sqrt{Y}$.
\end{enumerate}

These two pictures, although they endow most of the physical effects, are far from being totally understood. Among the improvements, one can quote phenomenological applications like geometric ad diffusive scaling predictions for vector-meson productions, for diffraction \cite{himst} and for LHC physics. On more theoretical grounds most of the work needed to be done concerns the effects of fluctuations. Most of the results known so far are derived within the local approximation for the noise term and neglecting the impact parameter dependence. In addition, apart from numerical studies, analytical results can only be applied to irrealistically small values of $\alpha_s$ and a better understanding for more physical values is still lacking. Although the existing picture is expected to contain the qualitative effects, a more precise treatment is certainly an interesting problem to address. 

Within the framework presented here, the link between the QCD evolution equations and equations from statistical physics has led to a large number of interesting ideas and results. Even in statistical physics, many questions are yet opened, hence, extending the QCD picture including more sophisticated treatments of the BFKL kernel and more precise forms of the noise term is certainly not a straightforward task. The questions of making clear predictions for phenomenology, of understanding the effects of fluctuations beyond the local-noise approximation and of including impact-parameter dependence are hence expected to give interesting work in the near future. Among particle physics, high-energy QCD is thus one of the most active fields. Because of those open questions and because of the requirement for an excellent knowledge of QCD at the LHC, we expect that high-energy QCD is still going to be active in the near future.

\begin{acknowledgements}
I would like to thank warmly the members of the Brazilian Particle Physics Society, and especially Beatriz Gay Ducati and Magno Machado, for the invitation to give a talk in this meeting. G.S. is funded by the National Funds for Scientific Research (FNRS), Belgium.
\end{acknowledgements}

\end{document}